\documentclass{article}
\begin{document}
\title{A short lecture on Divergences}
\author{A. Rivero \thanks{EU Politecnica de Teruel, Universidad de Zaragoza}}
\maketitle
\begin{abstract}
We present some clues to the study of the renormalization group, at graduate level,
as well as some bibliographical pointers to classical resources. Just the kind
of things one had liked to hear when starting to study the subject.
\end{abstract}

This was going to be a notice on ``recent advances on renormalization group theory'', from 
which advanced students can get bibliography to please their teachers. But I happened to 
visit Jacques Gabay's printing house, and I decided to take a wider view. At least, wider 
than usual treatises on Quantum Field Theory. Gabay's mission is to keep in print old 
mathematics texts from the late XIXth and early XXth, and his work helps to keep the 
perspective.

Fact is, all generations of physicists since Euler times have been used to live with 
divergences. Students are supposed to become exposed to the subject gradually, but this 
gradation varies strongly across schools and faculties, and the balance keeps more in the 
room of Cauchy than in Borel quarters. There is even a darker side, about if everything 
which is legal in Mathematics should be legal in Physics, but this question keeps usually in 
the philosophical level. Still, one should point that classical mechanics, the science of 
newtonian limits, is mathematically legal but physically ruled out!

The usual scenario for divergences is: we have a differential equation. We look for a 
solution from power series expansion. Most times, the solution 
is known to exist, say from Picard's fixed point method, say from other convergence 
theorem. But we are forced to pick up a power series expansion on the ``wrong'' parameter, 
so that the convergence radius is zero. Or (change $x-> 1/z$ if necesary) we could to know 
only a asymptotic series around infinity. Poincare treatises are the first ones showing all of 
this. 

Note, still, that we are here in a purely classical matter. We can do perturbation theory with 
a small coupling constant in a perturbation potential, and then expand the solution as a 
power series of the coupling constant. The series diverges, and then we have only an 
asymptotic series. Borel transform, or other resummation techniques, can be invoked to get 
a better expansion.  Sometimes even a convergent series is known, for instance for the 
three-body problem (which, remember, is chaotical respect to small perturbations), but even 
then it can happen that the divergent series is faster than the convergent one for a given 
level of precision!

This kind of problems when solving perturbation series is thus expected to appear 
also in quantum mechanics and quantum field theory. In fact,
an argument from Dyson \cite{dyson} shows that the perturbative expansion of quantum 
electrodynamics will give place to a divergent series of this kind. This is not to be confused 
with renormalizability questions, which will refer to each term in this expansion.

A interesting detail of the quantum approach is that it does not start from the
differential equation, but from the action principle. This is a puzzling thing, because the
action principle is an integral between two conditions, and it must be connected
to a local differential equation. It seems that quantum theory has an alternate
way to go from the lagrangian, in the action law, to the hamiltonian, in 
Schroedinger equation; it is very suggestive to read this derivation in the original 
article of Feynman\cite{feynman}. Later, Schwinger and Dyson produced a way to
deduce, from variation of a quantum action, the canonical equations of quantum theory.
In any case, calculation efficiency suggest to start always from perturbations
of the action, and this is the engine who drives to diagrammatics.

Also, we have a richer set of expansions when going to quantum theory. Besides the 
perturbative method around a coupling constant, we have the possibility to expand on Plank 
constant. In QM, the Schroedinger equation is written in Ricatti form,
$$i \hbar {d\xi \over dx} + 2 m (E-V) - \xi^2 =0 $$
and then we expand
$$\xi = \sum \xi_n(x)  (\hbar / i)^n$$
AFC \cite{AFC} refer to this as the eikonal expansion (This change of variables is 
characteristic also of the Hoft algebra of diffeomorfisms used by Connes and Kreimer).
And the diagrams give still other expansion; if we are in QFT with only a coupling constant,
then a series on Plank constant 
coincides with the expansion on number of loops. But for the general case, we have still this 
third possibility, on the number of loops of each Feynman diagram.

Besides perturbations, one could expect also divergences for bad shaped, singular, 
potentials, but their study has 
been bypassed by history. Still, it is worth to mention that the gravitating n-body problem 
can be formulated -and solved- including collisions throught the singularities. 

Well, this was the world of our grand-grandfathers in the first decades of XXth century. 
Then it came a more painful source of divergences. Not only happened that the perturbation 
expansion was around bad places, but also that every term in the expansion happened to 
contain a divergence of his own. As the effect relates to the indeterminate creation of 
virtual particles, this is the thing we are expected to learn in any QFT course. Still, it is 
worth to mention the three steps in taking control of this source of problems:

--First, Feynman, Dyson and Schwinger got to recognize how the divergence can be 
absorbed in the coupling constant in a consistent form. This comes from a lot of
previous experience on considering the cloud of virtual particles around a 
singular charge, and the real surprise is that just a finite number of coupling
constants can adjust for all the divergences.

--Second, Gellman and Low\cite{gl}, in a paper of compulsory reading, identify a general 
method to predict the renormalized values without entering in detailed calculation: the 
renormalization group. Asymptotical analysis get here a completely new meaning.

--Third, Kadanoff, Wilson and Kogut recognize the renormalization group as a question of 
scaling between different orders of magnitude. The study of fixed points, and perturbations
around them, conects this approach  with the previous results. 

In the first step, the problem is traced to an limit when $x-y$ goes to zero. In the Second 
step, the bug is transformed in a feature, and general properties of this $x-y$ divergence are 
systematised. In the third step, the question of existence of scales is related with the 
problem of continuum limit of a lattice. A renormalization ``semigroup'', decimating degrees 
of freedom, is sketched, and Wilson shows how it can be related to the previous G-L group. 
A fourth step could be the finding of asymptotically free theories; this is standard lore in 
your textbooks. 

Wilson's approach has the merit of making very explicit the role of the cut-off that
regularises the theory. In a renormalisable theory, this cut-off is removed after
the  manipulations and then the Gellman-Low equation (and Callan-Simanzik) become apparent.
It is possible to keep the cut-off and to do calculations with an ``effective'' theory,
even if it is not renormalisable. Feelings about this can be mixed. One paralell which
can be invoked is the origin of quantum mechanics: The Plank constand did appear in
physics as a mean to regularise the energy distribution, then getting finite results
where the Rayleigh-Jeans theory was divergent. Also other cut-off techniques are
natural in thermodinamics; in any case it is very delicate to extend these similtudes,
as one must distinguish between ultraviolet and infrared divergences; in this presentation
we are almost forgetting the IR ones.
 
After recognising the role of $x-y$ in the divergences, it is more natural to ask ``what about 
QM?'' in the straighforward formulation as $0+1$ dimensional field theory. Of course, power
 counting in a $0+1$ theory shows that any polynomial interaction 
will be free of divergences (in the sense of renormalization). But it is very interesting to 
study the role of $t_1-t_0$  from the point of view of scaling and renormalization theory. This 
exercise was published by Polonyi\cite{polonyi} in 1993\footnote{Tarrach in Barcelona, 
as well as Boya and myself in Zaragoza, did took some interest on different views of the 
renormalization process of QM}.

It is not only QM, by the way. The ideas coming from renormalization group, and perhaps 
the renormalization group itself, descent back to the theory of classical evolution of 
differential equations. There, dimensional analysis and scaling got a fresh airh from both 
Wilson and GellMann-Low techniques. We could mention the works of Goldenfeld-Chen-
Oono\cite{gco}, as well as the ones of Kunihiro, while AFC describes some previous 
attemps middle way between classical and quantum. Also the original question of 
resummation techniques (by changing the coefficients in the perturbative expansion) get 
some help from this, even if it  sounds strange, it seems it is possible to mimic QFT and to 
use the techniques of control of coefficients (the RG) to control the whole expansion.

In the mean time, since Wilson's age, the renormalization group had been rigorized from the 
point of view of mathematical physics. This task was accomplished by Bogoliugov, who 
worked out both the distributional character of the theory and the mechanics of the 
substraction process. The approach summarized into the BPHZ method, and additionaly 
some more analytical part of this research drove to the formulation of Epstein and Glaser. 
This was all the rigour a physicist could even desire.The surprise come recently when 
Connes and Kreimer found a rigorization from pure mathematics. Kreimer got a method to 
map Feynman diagrams to rooted trees, and it happened that this method formulated a Hoft 
algebra, nice enough to hold the group structure of renormalization. 

And this surprise came with a message from calculus, when it was noticed, by 
Brouder\cite{brouder}, that the Hoft algebra of trees was the same technique that had been 
used by Butcher in the seventies, to classify the numerical Runge Kutta methods! We can 
understand why if we look from a very general point of view, stratification of compactified 
configuration spaces, as Kreimer explains lately. Or we can go back in the time, when the 
series expansion of the solutions of a differential equation was classified by labelling each 
term with a Cayley  tree.

Start from a differential equation  $x'=F(x)$. One want to get the series
$x(t)= x_0 + x_1 t + x_2 t^2 + ...$, and it is straighforward that we can proceed
by derivating the original equation to get first
$$x''= F'(x) x'  =  F'(x) F(x)$$
then
$$x'''=F''(x) (x')^2 + F'(x) x'' = F'' F^2 + F' F' F $$
and so on. The increasing complexity of the expansion can be tamed by using rooted 
trees to describe each term\cite{conn.l}.

The same technique is used to tame the terms in a QFT expansion and to recursively
apply the renormalization, so it seems we are again approaching in the domain of
the theory of differential calculus. If the circle finally closes, we will have learn 
some very deep lessons on the structure of calculus and its interaction with physics.

\end{document}